\documentclass[conference,a4paper]{APSIPA2021}
\IEEEoverridecommandlockouts %
\usepackage{amsmath}
\usepackage{graphicx}
\usepackage{multirow}
\usepackage{threeparttable}
\usepackage[backend=biber,style=ieee,]{biblatex}
\usepackage{booktabs}
\usepackage{subfigure}
\usepackage{geometry}
\usepackage{fancyhdr}

\fancypagestyle{firststyle}{
  \fancyhf{}
  \fancyhead[C]{2023 Asia Pacific Signal and Information Processing Association Annual Summit and Conference (APSIPA ASC)}
}

\DeclareRobustCommand*{\IEEEauthorrefmark}[1]{%
	\raisebox{0pt}[0pt][0pt]{\textsuperscript{\footnotesize\ensuremath{#1}}}}

\begin{document}

\title{Enhanced Neural Beamformer with Spatial Information for Target Speech Extraction}

\author{
\authorblockN{
Aoqi Guo\authorrefmark{1}\thanks{Work conducted when the first author was intern at Xiaomi Inc.}, Junnan Wu\authorrefmark{2}, Peng Gao\authorrefmark{2}, Wenbo Zhu\authorrefmark{2}, Qinwen Guo\authorrefmark{1}, Dazhi Gao\authorrefmark{1,*}\thanks{* Corresponding author.} and Yujun Wang\authorrefmark{2}
}

\authorblockA{
\authorrefmark{1} Ocean University of China, China}

\authorblockA{
\authorrefmark{2} Xiaomi Inc., Beijing, China\\
E-mail: guoaoqi@stu.ouc.edu.cn dzgao@ouc.edu.cn}
}

\maketitle
\thispagestyle{firststyle}
\pagestyle{fancy}

\begin{abstract}
	Recently, deep learning-based beamforming algorithms have shown promising performance in target speech extraction tasks. However, most systems do not fully utilize spatial information. In this paper, we propose a target speech extraction network that utilizes spatial information to enhance the performance of neural beamformer. To achieve this, we first use the UNet-TCN structure to model input features and improve the estimation accuracy of the speech pre-separation module by avoiding information loss caused by direct dimensionality reduction in other models. Furthermore, we introduce a multi-head cross-attention mechanism that enhances the neural beamformer's perception of spatial information by making full use of the spatial information received by the array. Experimental results demonstrate that our approach, which incorporates a more reasonable target mask estimation network and a spatial information-based cross-attention mechanism into the neural beamformer, effectively improves speech separation performance.

\end{abstract}

\section{Introduction}
The target speech extraction task is derived from the speech separation task, which aims to extract the speech of a specific target speaker from a mixed signal. Unlike conventional speech separation, which deals with the permutation order of signals, target speech extraction only focuses on extracting the target signal, thus avoiding the permutation problem \cite{zmolikova2023neural}. In recent years, multi-channel beamforming algorithms have shown great potential in the task of target speech extraction. Beamforming algorithms achieve spatial domain filtering of the observed signal by enhancing the signal in the desired direction and suppressing the signal in other directions, which naturally meets the task goal of target speech extraction. Classical beamforming algorithms require accurate estimation of the wave direction and determination of the speech or noise part in the mixed signal \cite{gannot2017consolidated}, which is difficult to obtain in noisy and reverberant environments using traditional algorithms. Therefore, obtaining accurate target orientation information in real scenes, improving the estimation accuracy of the covariance matrix, and solving the numerical instability problem in matrix operations are the key to improving beamforming performance.

In recent years, the development of deep learning technology has provided new research ideas for solving the above problems. First, using neural networks for wave direction estimation has achieved good experimental results \cite{grumiaux2022survey}, providing more accurate target orientation information for beamforming algorithms. At the same time, combining neural networks with beamforming algorithms improves the estimation accuracy of the covariance matrix and effectively improves the performance upper limit of beamforming algorithms. Among them, Mask Based Beamforming, as a classic representative of the combination of deep learning and beamforming algorithms, uses a neural network to predict the mask of speech and noise to calculate the spatial covariance matrix of speech and noise, respectively, and then inputs it into traditional beamforming algorithms such as GEV or MVDR for numerical calculation \cite{erdogan2016improved}, \cite{higuchi2016robust}. For the numerical instability problem that traditional beamforming algorithms still have difficulty in solving even after diagonal loading \cite{li2003robust} in matrix numerical operations, ADL-MVDR \cite{zhang2021adl} used an RNN to replace the matrix inversion and eigendecomposition processes in the MVDR beamforming algorithm, avoided the numerical instability caused by singular matrices in matrix operations, and achieved significant performance improvement for the neural beamformer. In recent experiments, the combination of deep learning and beamforming algorithms has been further deepened. GRNNBF \cite{xu2021generalized} used an RNN to model the covariance matrices of speech and noise, replaced all matrix numerical operations in traditional beamforming algorithms, and directly predicted beamforming weights using neural networks, achieved good separation results. SARNN \cite{li2021mimo} added self-attention mechanism to the beamforming module, enhanced the neural network's modeling ability for the spatial and temporal information contained in the covariance matrix. EABNet \cite{li2022embedding} adopted a more aggressive approach, directly modelled the array observation signals to obtain higher-dimensional information representations than the covariance matrix. The neural network implicitly includes all the calculation steps of beamforming and directly outputs beamforming weights. The above methods combined with deep learning technology have all achieved good performance improvements for beamforming algorithms.

In scenarios such as in-car and remote conferences, the position of the target speaker relative to the microphone array is often fixed, and its angle information is easily obtained through the device \cite{zmolikova2023neural}. To leverage the angle information of the target sound source, Chen et al. proposed the wave direction feature calculated using the angle information of the target sound source, which enhanced the directionality of the speech separation system \cite{chen2018multi}. ADL-MVDR \cite{zhang2021adl}, GRNNBF \cite{xu2021generalized}, SARNN \cite{li2021mimo}, UFE \cite{wu2020end}, etc. inputted the angle feature into the neural network, improved the separation module's perception of spatial information, thereby improved the separation performance of the system. Gu et al. extended the angle feature from a two-dimensional plane to a three-dimensional space, improved the separation performance of the model when the azimuth angle between the target sound source and the interference source was small \cite{gu20213d}.

In this paper, we designed a model that uses target spatial orientation information to enhance the performance of neural beamforming. The system consists of a front-end pre-separation module and a back-end beamforming module. Compared with other models, our improvements mainly focus on the following two points:

$\bullet$
To improve the estimation accuracy of the spatial covariance matrices of speech and noise, we stack the input features in the channel dimension and use the UNet-TCN structure to model the input features. This avoids the loss of feature information caused by directly reducing the input features after connecting them in the frequency dimension.

$\bullet$
To better utilize the spatial information obtained by the array, we transform the spatial features of the input signal and the covariance matrix calculated by the pre-separation module into a feature space with the same dimensionality. We then use a cross-attention mechanism to enhance the beamforming network's perception of spatial orientation, thereby improving the prediction accuracy of the beamforming weights.

Experimental results show that after the improvements to the pre-separation module and beamforming module mentioned above, the target speech separation performance of the neural beamformer could be effectively improved.

The remainder of this paper is organized as follows. Section II presents the signal model for the target speech extraction task and the beamforming algorithm. Section III details our proposed neural beamforming algorithm. Section IV provides an overview of the experimental setup and presents an analysis of the experimental results. Finally, Section V concludes the paper.

\section{Signal Modeling and Beamforming}
\begin{figure}[t]
	\begin{center}
		\includegraphics[width=80mm]{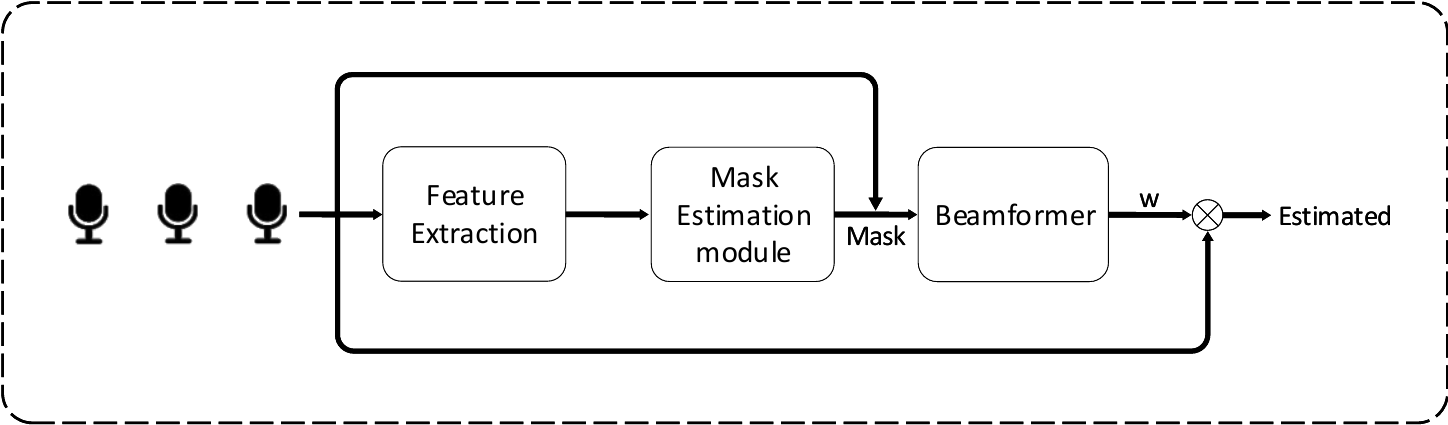}
	\end{center}
	\caption{General structure of a neural beamformer.}
	\vspace*{-3pt}
\end{figure}
Consider the far-field time-domain signal model of the real scene, described as:
\begin{equation}
	y(t) = x(t)*a_1(t) + n(t)*a_2(t) + s(t)
\end{equation}
\begin{equation}
	= x^{'}(t) + n^{'}(t) + s(t)
\end{equation}
where $ y(t) = [y^{(0)}(t), y^{(1)}(t), ..., y^{(M-1)}(t)]^T $ indicating the time-domain signal received by the M-channel microphone array. $x(t), n(t)$ represent the reverberation-free speech signals from the target speaker and the interfering speaker, respectively. $a_1(t), a_2(t)$ are the room impulse response(RIR) between the speaker and the array elements of the microphone. $*$ denotes the convolution operation. $x^{'}(t), n^{'}(t)$ represent the speech signals of the target speaker and the interfering speaker with reverberation, respectively. $s(t)$ is the background noise. When not focusing on the dereverberation task, the task goal of target speech extraction is to extract the target speaker's speech $x^{'}(t)$ from the noisy signal $y(t)$. 

After the signal model is transferred to the time-frequency domain by short-time fourier transform(STFT), it is expressed as:
\begin{equation}
	Y(t,f) = X(t,f) + N(t,f) + S(t,f)
\end{equation}

The purpose of the beamforming algorithm is to obtain the filter weight $w$ for the array observation signal, and extract the desired signal from the array observation signal by performing spatial filtering on the observation signal, that is:
\begin{equation}
	X(t,f) = w^H Y(t,f)
\end{equation}
where $(\cdot)^H$ denotes the conjugate transpose operation.

Taking the traditional MVDR beamforming algorithm as an example, the purpose is to minimize the output noise power of the beamforming algorithm without distorting the signal, that is:
\begin{equation}
	min(w^H \Phi_{NN} w^H),\quad s.t\quad w^H \alpha(\theta) = 1
\end{equation}

For the above formula, the desired filter weight $w$ is solved by the lagrange multiplier method:
\begin{equation}
w = \frac{\Phi_{NN}^{-1} \alpha(\theta)}{\alpha^H(\theta) \Phi_{NN}^{-1} \alpha(\theta)}
\end{equation}
where $\Phi_{NN} = N(t,f)N^H(t,f), \Phi_{NN}\in M\times M $ is the covariance matrix of the noise signal, $\alpha(\theta) = [e^{-j\omega\tau_m}], m = [0, 1, ..., M-1]$ is the steering vector that calculated using target sound source's DOA. When the steering vector is difficult to obtain due to the interference of environmental factors, PCA can be performed on the covariance matrix of the speech signal to obtain the estimated value of the steering vector \cite{souden2009optimal}.
\begin{figure*}[t]
	\begin{center}
		\includegraphics[width=180mm]{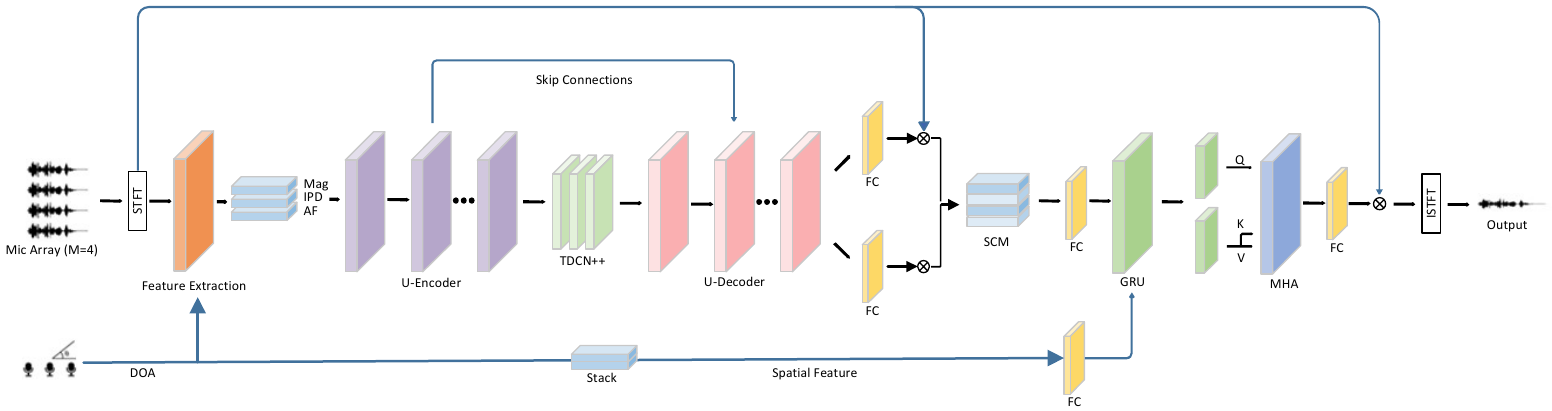}
	\end{center}
	\caption{The overall structure of our proposed model.}
	\vspace*{-3pt}
	{\hfill\footnotesize Both the input features and the covariance matrix are stacked along the channel dimension when inputting to the network.\hfill}
\end{figure*}

\begin{figure}[t]
	\begin{center}
		\includegraphics[width=70mm]{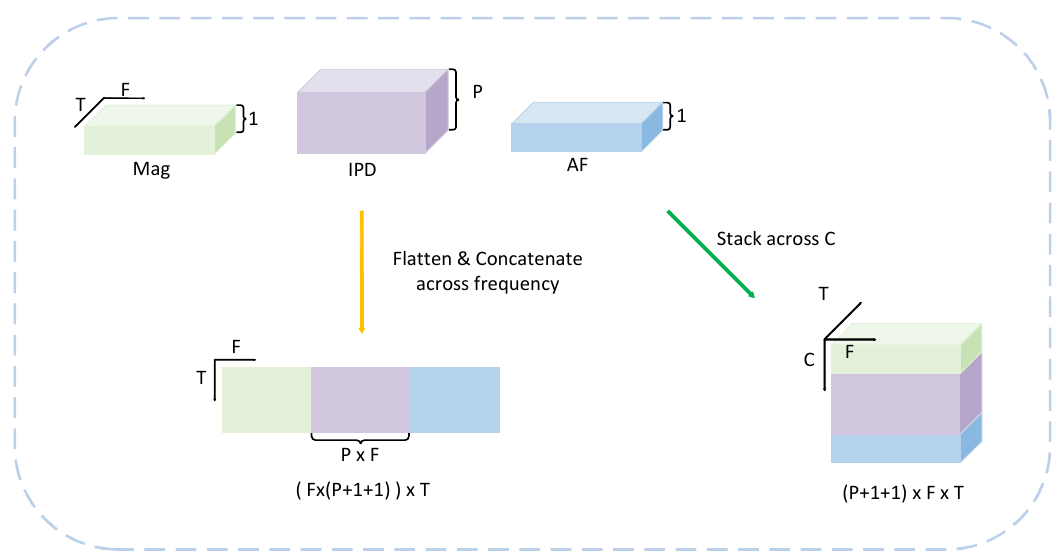}
	\end{center}
	\caption{The combination of input features.}
	\vspace*{-3pt}
\end{figure}

\begin{figure}[t]
	\begin{center}
		\includegraphics[width=72mm]{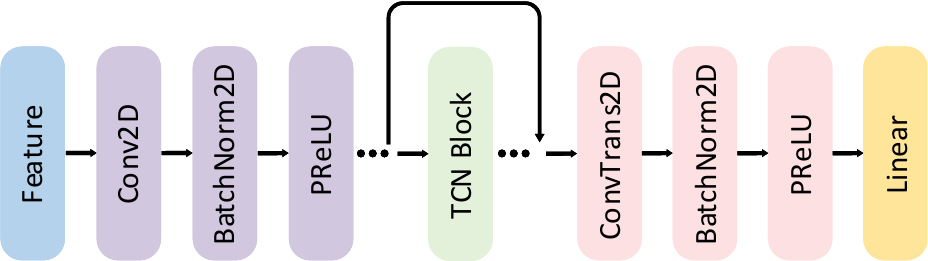}
	\end{center}
	\caption{Pre-separator based UNet-TCN structure}
	\vspace*{-3pt}
\end{figure}

Therefore, it can be seen that the key to the beamforming algorithm lies in the estimation of the covariance matrix. Traditional beamforming methods often rely on certain algorithms to determine the time periods of speech and noise in the signal, but these approach is not optimal. With the development of deep learning technology, using deep neural networks to determine the speech and noise components in the signal has achieved good experimental results.

Fig. 1 shows the general framework of the combination of neural network and beamforming. It first predicts a set of time-frequency masks representing the corresponding relationship between the desired signal and the original mixed signal through the neural network, such as IBM \cite{erdogan2016improved}, IRM \cite{higuchi2016robust}, CRM \cite{williamson2015complex}, CRF \cite{schroter2022deepfilternet}, etc. Taking the IRM as an example, it defines the energy ratio relationship between the desired signal and the mixed signal:
\begin{equation}
	IRM_x = \frac{|X(t,f)|}{|Y(t,f)|}
\end{equation}

\begin{figure}[t]
	\begin{center}
		\includegraphics[width=72mm]{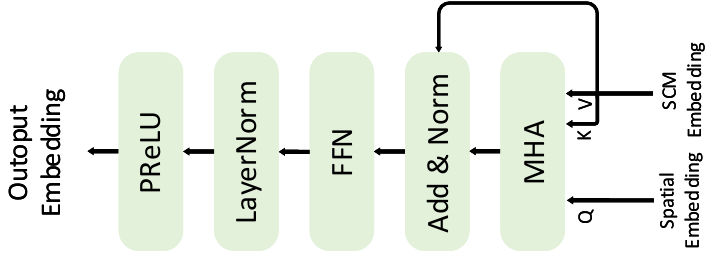}
	\end{center}
	\caption{A cross attention module with spatial information as Query and covariance matrix information as Key and Value. All features are on the time frame level.}
	\vspace*{-3pt}
\end{figure}

Then the speech and noise components in the mixed signal can be recovered through mask, so as to calculate the speech and noise space covariance matrix respectively \cite{erdogan2016improved}:
\begin{equation}
	\Phi = \frac{\sum_{t=1}^{T}M^2(t,f)Y(t,f)Y^H(t,f)}{\sum_{t=1}^{T}M^2(t,f)}
\end{equation}

Once the covariance matrix of the target signal and noise is calculated, it can be substituted into traditional beamforming algorithms such as GEV and MVDR to obtain the beamforming weight $w$. Recent experiments have demonstrated that utilizing neural networks to model covariance matrix information and predict beam weights can lead to improved target speech separation compared to traditional algorithms.

\section{Enhanced Neural Beamformer with Spatial Information}
Fig. 2 shows the overall architecture of our proposed model. First, the observation signal of the array is transformed into the time-frequency domain by STFT to extract the spatial features of the signal. The pre-separation network then estimates the mask of the target speech and noise based on the input features and calculates the covariance matrix. Following this, the beamforming network leverages the spatial features to predict the beamforming weight, which is subsequently used to beamform the observed signal. Finally, the original time domain signal is restored through ISTFT. Specifically as follows:
\subsection{Feature Extraction}
The first microphone from the multi-channel mixed signal is selected as the reference microphone, and its received signal's amplitude spectrum is calculated as the input feature for the pre-separation network.
\begin{equation}
	Y^{mag}_1(t,f) = |Y_1(t,f)|
\end{equation}
To obtain the spatial information of the microphone array, we select P pairs of microphones and calculate the phase difference between each pair as the spatial feature for the input of our model. This approach has been widely used in speech enhancement and source separation tasks, as it provides a concise and effective way to capture the spatial information of the microphone array.
\begin{equation}
	IPD_{i,j}(t,f) = \angle Y_i(t,f)-\angle Y_j(t,f)
\end{equation}
$$
i,j\in (0,1,...,M-1), i\neq j
$$
In order to enhance the spatial information acquired by the model, the cosine distance between the phase delay {\bf $\varphi_1$} of the signal in the desired direction on the array and the phase difference $\varphi_2$ observed by a certain pair of microphones is calculated. Then sum over all P pairs of microphones. That is, the angular characteristics of the target signal serve as supplementary input.
\begin{equation}
	AF(\theta,f) = \sum_{p=1}^P cos(IPD_{i,j}(t,f)-\frac{2\pi fdcos\theta}{c})
\end{equation}
When the phase angle of the mixed signal and the desired signal is closer, the value of the angle characteristic tends to 1 approximately \cite{wu2020end}. All these features are stacked along the channel dimension to serve as the network's input features.
$$Feature = [Y^{mag}_1(t,f), cosIPD_{i,j}(t,f), AF(\theta,f)]$$

\subsection{Pre-separation Module}

The pre-separation network adopts an Encoder-TCN-Decoder structure, which includes 2D convolutional layers, linear layers, and a variant of TCN \cite{wisdom2020unsupervised}. To process the input features, it's a common practice to concatenate them on the frequency dimension, followed by 1D convolution to reduce dimensionality before feeding them into the TCN or RNN module \cite{zhang2021adl}, \cite{xu2021generalized}, \cite{li2021mimo}. However, to avoid the loss of feature information due to direct dimensionality reduction of the input features, we employ the UNet-TCN structure to model them. By stacking the input features along the channel dimension, we obtain $Y_{input}\in R^{N\times F\times T} (N=P+1+1)$, which is then fed into the UNet-TCN layer to model the nonlinear mapping of the time dimension from the source signal domain to the separated signal domain and noise domain. Finally, the predicted cRM is restored through the linear layer to recover the target speech.
\begin{equation}
	X_{cRM}(t,f) = (cRM_r+jcRM_i)(Y_r+jY_i)
\end{equation}
The noise signal is also calculated in the above way. All activation functions in the pre-separation network use PReLU to accelerate the convergence speed of the network. Compared with ReLU, it allows negative values to appear, which is synergistic with the negative values between the input phase difference and the output mask.

\subsection{Neural Beamforming Module}
The beamforming network is composed of a recurrent neural network and a cross-attention module. First, the speech and noise signals generated by the pre-separation of the front-end network are used to calculate the spatial covariance matrix $\Phi_{SS}(t,f)\in C^{M\times M}, \Phi_{NN}(t,f)\in C^{M\times M}$ at the time frame level respectively.
\begin{equation}
	\Phi_{SS}(t,f) = X_{cRM}(t,f) X_{cRM}^{H}(t,f) 
\end{equation}
\begin{equation}
	\Phi_{NN}(t,f) = N_{cRM}(t,f) N_{cRM}^{H}(t,f) 
\end{equation}
Since the spatial covariance matrix is a complex-valued Hermitian matrix, and the defined neural network is a real network, the real and imaginary parts of the speech and noise signal covariance matrices are concatenated on the channel dimension, and LayerNorm is used to the covariance matrix of is standardized. In order to enhance the spatial constraints on the beamformer, we use the spatial information received by the array, that is, the phase difference between channels and the angle characteristics of the target sound source, to perform a cross attention mechanism on the covariance matrix of speech and noise. First, the linear layer and GRU are used to transform the spliced covariance matrix and cosIPD, Angle-Feature(AF) to model the inter-channel information at the time frame level and transform it into embedding of the same dimension. Subsequently, the modeled spatial information is used as the query of multi-head attention, and the covariance matrix information is used as the key and value, and the spatial orientation information is used to enhance the ability of attention to model inter-channel information at the time frame level.
\begin{equation}
	Query = RNN-DNN(cosIPD, AF)
\end{equation}
\begin{equation}
	Key, Value = RNN-DNN(\Phi_{SS}, \Phi_{NN})
\end{equation}
\begin{equation}
	Attention(Q,K,V) = softmax(\frac{QK^T}{\sqrt{d_k}})V
\end{equation}

Then, the observed signals from the microphones are beamformed using the beamforming weights output by the linear layer to obtain the spectrogram of the model prediction output. The time-domain signal is then recovered through ISTFT, resulting in the desired time-domain speech signal.

\section{Experiment setup and result analysis}
\subsection{Dataset settings}
The source speech signals for the target and interfering speakers were obtained from the Aishell2 \cite{du2018aishell} dataset, while the background noise data was sourced from DNS2023 \cite{dubey2023icassp}. For an indoor reverberation scene, the room size was set to a random size between [3, 3, 1.5] and [8, 8, 2.5] meters for length, width, and height, and the reverberation time(rt60) was set to 0.1-0.6 seconds. The simulation was carried out using the image source method, generating room impulse responses for the target and interference signals \cite{scheibler2018pyroomacoustics}. The microphone array consisted of a four-element linear array with an element spacing of 3 cm. To ensure a certain degree of spatial independence in the signals, we set the minimum angle between the two sources relative to the microphone array to 5°. The signal-to-noise ratio between the target and interference signals was set to [-6, 6] db. To enhance the model's robustness, a background noise of [-5, 20] db was added to the simulated data. In total, the simulation generated 120,000 pieces of training set data, 14,000 pieces of validation set data, and 7,000 pieces of test set data. All audio was down-sampled to 16kHz and each piece of data is 4 seconds long.

\subsection{Experiment settings}
During the training process, the window length of STFT and the number of FFT points are set to 32ms, and the frame shift is set to 16ms. For the array, we select (0,1), (0,2), (0,3) three microphone pairs to calculate cosIPD. Therefore, the channel dimension of the input feature is 5, that is $Feature \in R^{5\times F\times T}$. For the pre-separation module, the number of channels of Conv2D in the UNet structure is (5, 32, 64, 128). The TCN structure is a 3x8 stacked TCN Block with 128 channels. For the beamforming module, the number of hidden layer units of the double-layer GRU network is 256, the dimension of the cross attention module is 128, and the last linear layer outputs complex-valued beamforming weights, so the output dimension is set to 8 (channels*2), the proposed The model is trained end-to-end.

The experiment uses two A100 graphics cards to train 60 epochs on the simulation data set, and the Batch Size is set to 24. The Adam optimizer is used for optimization, the initial learning rate is set to 2e-3 and decays continuously with the number of epochs, and the decay coefficient is 0.98. The maximum gradient clipping is set to 10 to speed up the network convergence process.

We use the MVDR method based on the IRM and the GRNNBF \cite{xu2021generalized} and SARNN\cite{li2021mimo} models trained according to the MISO model as the baseline for experimental comparison. For IRM-MVDR, we use the TCN network to predict the IRM, which is then substituted into the MVDR algorithm for beamforming. For GRNNBF and SARNN, the 4x8 TCN Block is used to predict the CRF mask, which is input to the GRU or self-attention module with 500 hidden units after LayerNorm to output beamforming weights. For all models, we use the joint loss function of SI-SDR and MSE for training, and the weights of the two are equal.

\subsection{Results}

\begin{table}[t]
	\begin{center}
	\caption{SI-SDR(DB), PESQ AND STOI RESULTS OF IRM-MVDR, GRNNBF, SARNN AND OUR PROPOSED MODELS.}
	\begin{tabular}{p{7em}cccc}
		\toprule
		\centering Model & \multicolumn{1}{p{4.69em}}{\centering Para.(M)} & \multicolumn{1}{p{4.19em}}{\centering PESQ{\bf↑}} & \multicolumn{1}{p{4.19em}}{\centering STOI{\bf↑}} & \multicolumn{1}{p{4.25em}}{\centering SISDR{\bf↑}} \\
		\midrule
		\centering No processing & \multicolumn{1}{p{4.69em}}{\centering -} & 1.148 & 0.563 & -1.76 \\
		\centering IRM MVDR{\bf\cite{higuchi2016robust}} & 5.33  & 1.586 & 0.757 & 5.25 \\
		\centering GRNNBF{\bf\cite{xu2021generalized}} & 15.73 & 2.176 & 0.845 & 8.42 \\
		\centering SARNN{\bf\cite{li2021mimo}} & 16.99 & 2.324 & 0.859 & 9.48 \\
		\centering Proposed & 8.64  & {\bf2.499}   & {\bf0.88}  & {\bf10.51} \\
		\bottomrule
	\end{tabular}%
	\label{tab:addlabel}%
	\end{center}
\end{table}%

\begin{table}[t]
	\centering
	\caption{SI-SDR(DB), PESQ AND STOI RESULTS OF ABLATION EXPERIMENTS WITH OUR PROPOSED MODEL}
	\begin{tabular}{p{9.00em}cccc}
		\toprule
		\centering Model & \multicolumn{1}{p{4.69em}}{\centering Para.(M)} & \multicolumn{1}{p{4.19em}}{\centering PESQ{\bf↑}} & \multicolumn{1}{p{4.19em}}{\centering STOI{\bf↑}} & \multicolumn{1}{p{4.25em}}{\centering SISDR{\bf↑}} \\
		\midrule
		\centering TCN IRM (cat) & 5.33  & 1.586 & 0.755 & 5.95 \\
		\centering + UNet IRM (stack) & 6.76  & 1.846 & 0.799 & 6.82 \\
		\centering TCN cRM (cat) & 5.4   & 1.591 & 0.759 & 5.97 \\
		\centering + UNet cRM (stack) & 7.09  & 1.851 & 0.801 & 6.86 \\
		\centering + BF(MHSA 128) & 8.05  & 2.466 & 0.876 & 10.11 \\
		\centering + BF(MHSA 256) & 8.89  & 2.481 & 0.878 & 10.41 \\
		\centering + BF(MHCA) & 8.64  & {\bf2.499}   & {\bf0.88}  & {\bf10.51} \\
		\bottomrule
	\end{tabular}%
	\label{tab:addlabel}%
\end{table}%

\begin{figure}[htbp]
	\subfigure[Noisy, SI-SDR: -3.4dB]{
		\centering
		\includegraphics[width=0.24\textwidth]{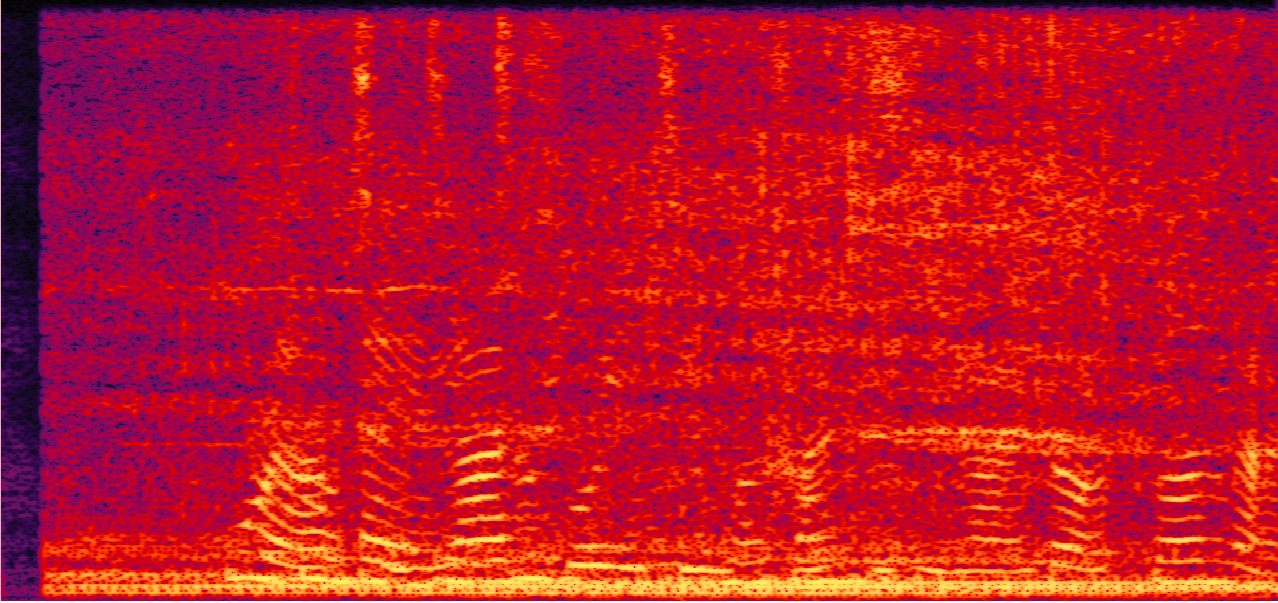}
	}\subfigure[IRM-MVDR, SI-SDR: 5.4dB]{
		\centering
		\includegraphics[width=0.24\textwidth]{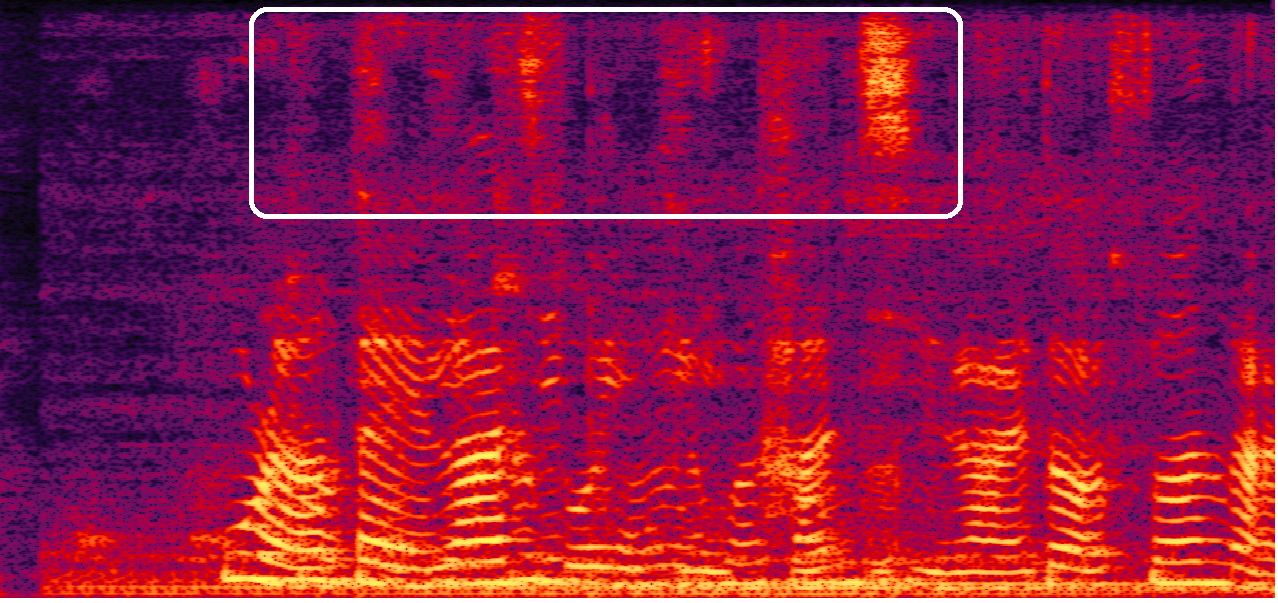}
	}
	
	\subfigure[GRNNBF, SI-SDR: 6.91dB]{
		\centering
		\includegraphics[width=0.24\textwidth]{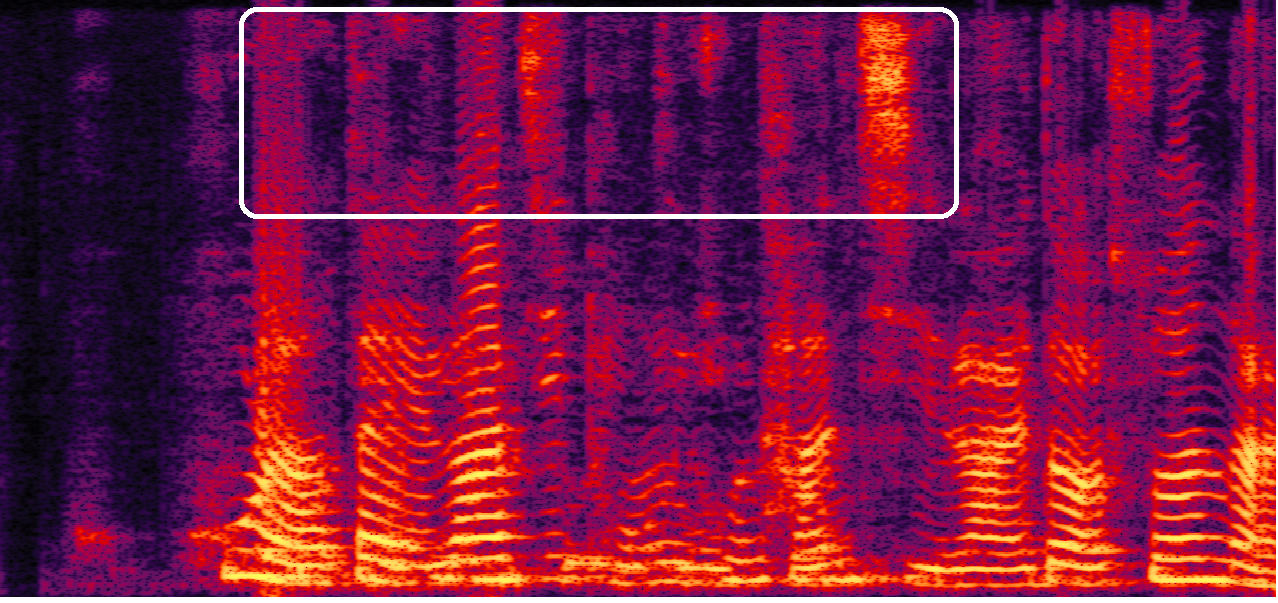}
	}\subfigure[SARNN, SI-SDR: 7.81dB]{
		\centering
		\includegraphics[width=0.24\textwidth]{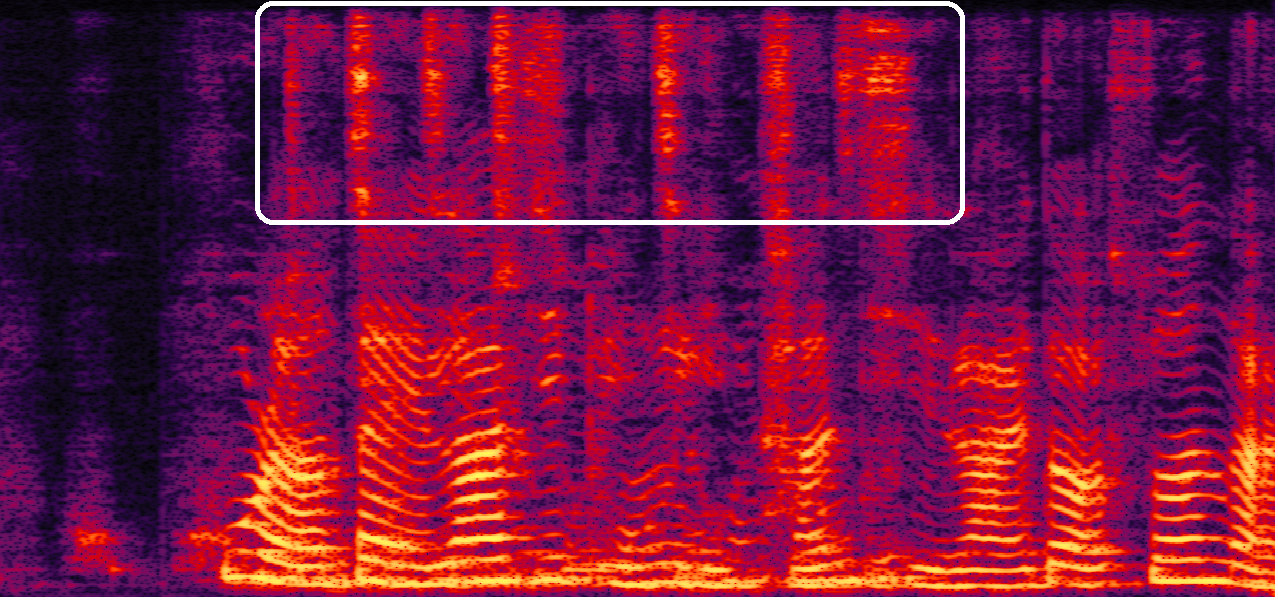}
	}
	
	\subfigure[Proposed, SI-SDR: 9.97dB]{
		\centering
		\includegraphics[width=0.24\textwidth]{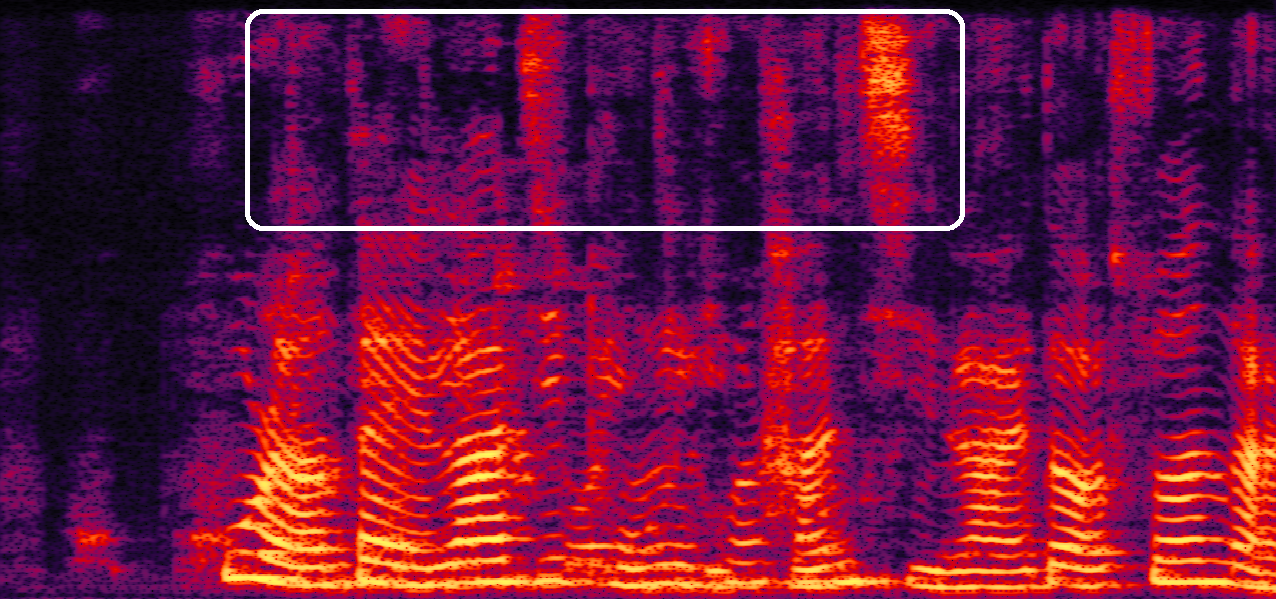}
	}\subfigure[Target]{
		\centering
		\includegraphics[width=0.24\textwidth]{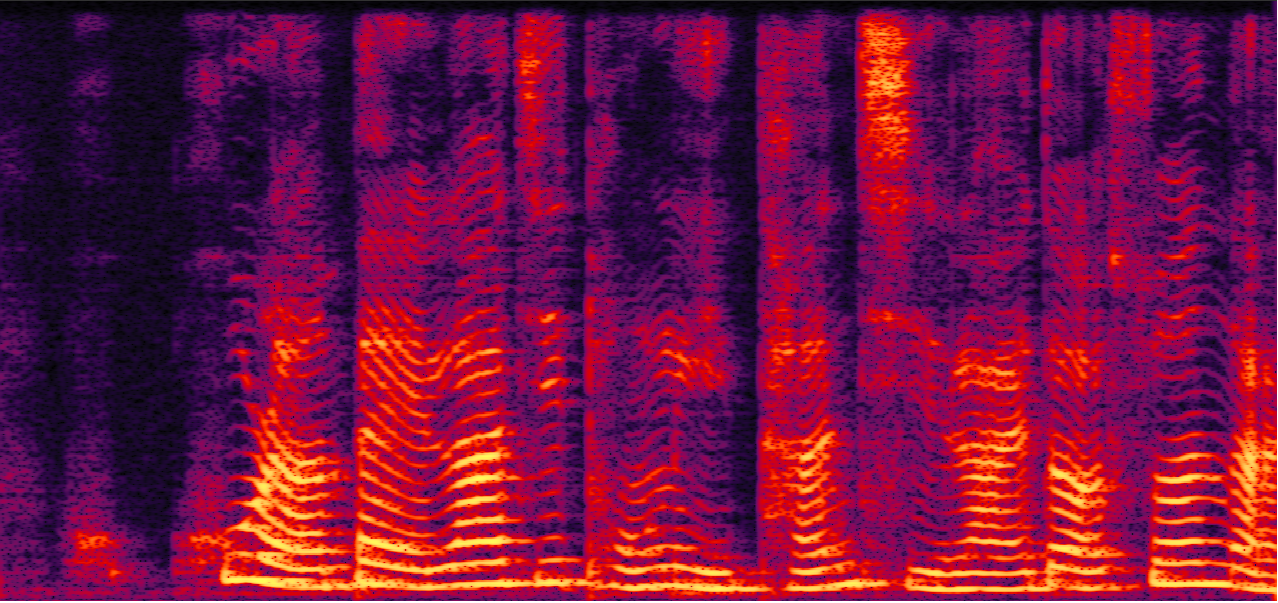}
	}\caption{Spectrum of a piece of data in the test set after model processing. SIR=-2.2dB, SNR=-4.8dB. The angle between the speaker is 9°.\label{inut_output}}
\end{figure}

Table 1 shows the experimental results of our proposed model and baseline model on the test set. In terms of model parameters, our model has increased compared with IRM-MVDR, which is mainly due to the difference in the prediction mask and the replacement of the MVDR algorithm with a neural beamforming module. Compared with GRNNBF and SARNN, our parameter reduction is more obvious. This is because after adding the cross-attention mechanism based on spatial information, the model does not require a large number of hidden layer units to achieve a satisfactory separation effect. Therefore, the experimental results show that our proposed model can significantly improve the separation performance of the target speech compared with the baseline while reducing the number of parameters.

In Table 2, we present the results of our ablation experiments on the proposed model. The experimental results show that for the pre-separation module, the pre-separation effect obtained by first encoding the input features through the Unet structure and then inputting them to the TCN block to output the mask is better than directly inputting the input features to the TCN through Conv1D. For the neural beamformer, the UNet-TCN structure makes the model more fully utilize the input feature information, reduces the information loss in the dimension reduction process, and makes the estimation of the covariance matrix more accurate. In terms of mask prediction, whether it is TCN or UNet-TCN structure, the effect of network prediction cRM on recovering speech signals is slightly better than that of IRM, but the gap is not large. This is because for frequency-domain speech separation, when the frame length is 32ms, the effect of phase estimation on the separation effect is limited \cite{peer2022phase}. 

For the neural beamforming module, we take multi-head self-attention(MHSA) module with 128 and 256 hidden layer units as an example. It is evident that incorporating the embedding modeled by the covariance matrix and spatial information into the multi-head cross-attention(MHCA) module leads to better speech separation performance than the MHSA module with an equivalent number of hidden layers. Experimental results demonstrate the effectiveness of introducing spatial information into the beamforming module via the cross-attention mechanism.

\subsection{Spectrogram analysis}
We analyze and compare the spectrograms of the target speech generated using different separation methods, and discuss the results further. In order to verify the robustness of the algorithm, we selected a piece of data with a low SNR and a small angle between the target speaker and the interference speaker from the test set for visualization, as shown in Fig. 6. Due to numerical instability in the training process and the influence of the MVDR algorithm itself, the IRM-MVDR model has a poor suppression effect on interference noise, and the separated spectrum is severely damaged by noise. In GRNNBF and SARNN using neural beamers, the separation performance of the system has been improved to a certain extent, and both have a good suppression effect on background noise, but there may still be some residual interference noise present in the separated signal. After the proposed model uses the UNet-TCN structure to improve the pre-separation network and adds spatial information to the beamforming module, the background and interference noise are further suppressed, and the degree of spectral distortion is lower than that of the baseline model, achieved better experimental results.

\subsection{Beam pattern}
To further analyze the spatial filtering performance of the model, Fig. 7 visualizes the beam patterns of different neural beamformers, including GRNNBF, SARNN and our proposed model. The speech is divided into four segments, segment (a) has only background noise, segment (b) has two speakers speaking at the same time, segment (c) has only the interfering speaker active, segment (d) has only the target speaker active. All beam patterns are averaged over the frequency dimension.

All three models have a strong suppression near 90° when there is no sound source activity, which is a little strange, we don't know if this is due to the influence of background noise. When two sound sources speak at the same time, all three models produce a strong suppression effect on the interference direction. Our proposed model suppresses the disturber to a greater extent when only the disturber is active. When only the target sound source is active, the directivity of our proposed model is better than that of GRNNBF and SARNN, which reduces the degree of distortion of the model-predicted spectrum.

\section{Conclusion}
In this study, we propose a novel UNet-TCN structure to model input features, resulting in improved estimation accuracy of the covariance matrix. Furthermore, we introduce spatial information through the cross-attention module in the neural beamforming module to enhance beamformer performance, enabling better noise rejection and reduced spectral distortion. Objective analysis and subjective evaluation show significant improvements in separation effect. In the future, we plan to achieve better target speech extraction while reducing model complexity and output spectrum distortion.

\begin{figure}[t]
	\begin{center}
		\includegraphics[width=90mm]{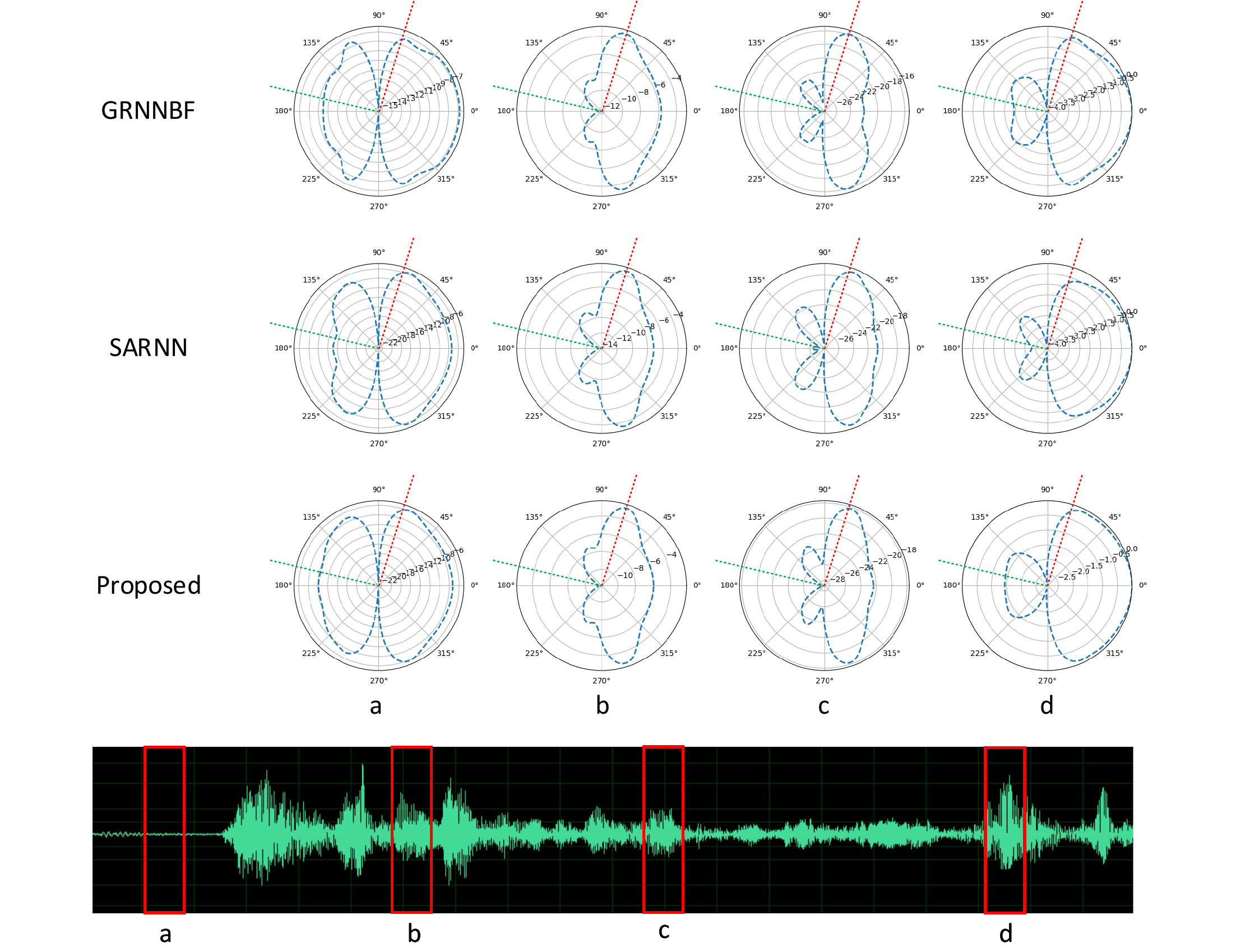}
	\end{center}
	\caption{An example of beam patterns derived by GRNNBF, SARNN and our proposed module. The DOA of the target speaker and the interference speaker are 79 and 164, and marked with red and green respectively.(a) is a silent segment, (b) is two speakers are active at the same time, (c) is only the interference speaker, (d) is only the target speaker.}
	\vspace*{-3pt}
\end{figure}
\newpage
\printbibliography

\end{document}